\documentclass[12pt,preprint]{aastex}
\shorttitle{The r-Process}
\shortauthors{Terasawa et al.}

\begin{document}

%% LaTeX will automatically break titles if they run longer than
%% one line. However, you may use \\ to force a line break if
%% you desire.

\title{r-Process Nucleosynthesis in Neutrino-Driven Winds from a Typical
Neutron Star with $M = 1.4 M_{\odot}$}

%% Use \author, \affil, and the \and command to format
%% author and affiliation information.
%% Note that \email has replaced the old \authoremail command
%% from AASTeX v4.0. You can use \email to mark an email address
%% anywhere in the paper, not just in the front matter.
%% As in the title, you can use \\ to force line breaks.

\author{M.\ Terasawa\altaffilmark{1}, K.\ Sumiyoshi\altaffilmark{2}, S.\
Yamada\altaffilmark{3}, H. Suzuki\altaffilmark{4}, and T.\
Kajino\altaffilmark{1}}

%% Notice that each of these authors has alternate affiliations, which
%% are identified by the \altaffilmark after each name.  Specify alternate
%% affiliation information with \altaffiltext, with one command per each
%% affiliation.

%\altaffiltext{1}{Department of Astronomy, School of Science, University
%of Tokyo, Hongo, Bunkyo-ku, Tokyo 113-0033, Japan}
\altaffiltext{1}{National Astronomical Observatory, 
Osawa, Mitaka, Tokyo 181-8588, Japan}
\altaffiltext{2}{Numazu College of Technology, Ooka, Numazu, Shizuoka
410-8501, Japan}
\altaffiltext{3}{Institute of Laser Engineering (ILE), Osaka University,
Yamadaoka, Suita, Osaka 565-0871, Japan}
\altaffiltext{4}{Faculty of Science and Technology, Tokyo University of Science,
Yamazaki, Noda, Chiba 278-8510, Japan}

%% Mark off your abstract in the ``abstract'' environment. In the manuscript
%% style, abstract will output a Received/Accepted line after the
%% title and affiliation information. No date will appear since the author
%% does not have this information. The dates will be filled in by the
%% editorial office after submission.

\begin{abstract}

We study the effects of the outer boundary conditions in
neutrino-driven winds on the r-process nucleosynthesis.  We perform
numerical simulations of hydrodynamics of neutrino-driven winds and
nuclear reaction network calculations of the r-process.  As an outer
boundary condition of hydrodynamic calculations, we set a pressure
upon the outermost layer of the wind, which is approaching toward the
shock wall.  Varying the boundary pressure, we obtain various
asymptotic thermal temperature of expanding material in the
neutrino-driven winds for resulting nucleosynthesis.  We find that the
asymptotic temperature slightly lower than those used in the previous
studies of the neutrino-driven winds can lead to a successful
r-process abundance pattern, which is in a reasonable agreement with
the solar system r-process abundance pattern even for the typical
proto-neutron star mass $M_{NS} \sim 1.4 M_{\odot}$.  A slightly lower
asymptotic temperature reduces the charged particle reaction rates and
the resulting amount of seed elements and lead to a high
neutron-to-seed ratio for successful r-process.  This is a new idea
which is different from the previous models of neutrino-driven winds
from very massive ($M_{NS} \sim 2.0$ $M_{\odot}$) and compact ($R_{NS}
\sim 10$ km) neutron star to get a short expansion time and a high
entropy for a successful r-process abundance pattern.  Although
such a large mass is sometimes criticized from observational facts on
a neutron star mass, we dissolve this criticism by reconsidering the
boundary condition of the wind.  We also explore the relation between
the boundary condition and neutron star mass, which is related to the
progenitor mass, for successful r-process.

\end{abstract}

%% Keywords should appear after the \end{abstract} command. The uncommented
%% example has been keyed in ApJ style. See the instructions to authors
%% for the journal to which you are submitting your paper to determine
%% what keyword punctuation is appropriate.

\keywords{r-process nucleosynthesis, supernova, neutron star}

%% From the front matter, we move on to the body of the paper.
%% In the first two sections, notice the use of the natbib \citep
%% and \citet commands to identify citations.  The citations are
%% tied to the reference list via symbolic KEYs. The KEY corresponds
%% to the KEY in the \bibitem in the reference list below. We have
%% chosen the first three characters of the first author's name plus
%% the last two numeral of the year of publication as our KEY for
%% each reference.

\section{Introduction}

The r-process nucleosynthesis, which is a rapid neutron capture process
faster than beta-decay, is believed to be responsible for about a half
of elements heavier than iron (Burbidge et al. 1957).  Many heavy
elements presumably produced in the r-process have recently been
detected in extremely metal-poor stars by recent astronomical
observations (Sneden et al. 1996, 1998, 2000, Westin et al. 2000,
Johnson and Bolte 2001, Cayrel et al. 2001, Honda 2001).  Their
abundance pattern proves to be very similar to the one of the solar
r-process pattern, which is called the universality in the r-process
abundance pattern, especially in the region of $56 \le Z \le 70$.  This
universality of abundance pattern strongly suggests that the r-process
occurs in the same way independently of the metallicity along the entire
history of Galactic chemical evolution from the beginning of the Galaxy
to the present. It means simultaneously the fact that the origin of the
r-process is most likely in  supernovae (SNe) of
massive progenitor stars since massive stars first evolve and end up
with SN explosions whose ejecta reflects  abundances in metal-poor stars.

One of the most plausible sites of the r-process is the
neutrino-driven winds in supernovae.  Woosley et al. (1994) have
demonstrated that the very high entropy conditions, $\sim 400$ k$_B$,
are realized in the neutrino-driven wind and on these specific conditions
the r-process nucleosynthesis occurs successfully.  This r-process
scenario in high entropy hot bubble has been later pointed out to be
rather difficult because copious supernova neutrinos change neutrons
to protons and hinder the r-process (Meyer 1995).  In more recent
studies, the successful r-process pattern in the neutrino-driven winds
has been obtained even for relatively low entropy, $\sim 200$k$_B$,
provided that the expansion time scale is much shorter, $\sim 10$ ms,
than the time scale of the neutrino process (Otsuki et al. 2000,
Sumiyoshi et al. 2000), making the neutrino-driven wind scenario
viable again.  However, they assumed a large neutron star mass, $\sim
2.0 M_{\odot}$, in order to gain a slightly higher entropy.  These
parameter setups have been referred with caution because the observed
neutron star has a typical mass $\sim 1.4 M_{\odot}$ and a radius
$\sim 10$ km.

In this paper we will discuss that the outer boundary conditions of
the neutrino-driven winds may resolve this problem of the neutron star
mass. It has recently been founded (Terasawa et al. 2001) that the
r-process occurs far from the neutron star in a less dense region
rather close to and behind the outward shockwave after the surface
material is blown off.  It is generally known that alpha captures are
very sensitive to the temperature. A little change in temperature can
make a large effect on the abundance of seed elements and heavy
r-process elements. This effect on nucleosynthesis of outer region has
not been studied very well.  In most previous studies, the boundary
condition of outer region has been chosen based on the results by
Woosley et al. (1994). This is because their simulation is the unique
simulation which deals with both supernova explosion and
neutrino-driven wind at once. In their study the temperature is about
0.1 MeV and the density is $\sim 10^{3}$ g/cm$^3$ at the outer
boundary (at radius of 10$^{4}$ km).

The physical condition in the outer boundary region depends on the
competition between the falling matter from the envelope and the
outward shockwave, giving many factors to change by a core bounce.
There are a lot of factors to effect the boundary condition.  We
perform the hydrodynamical simulations of neutrino-driven winds and
investigate the dependence of the r-process nucleosynthesis on the
pressure of the outer material, toward which the wind blows, behind
the shockwave.  We show that the successful r-process can occur in the
neutrino-driven winds from a typical neutron star mass with $1.4
M_{\odot}$ by choosing a slightly lower pressure and temperature than
those adopted previously.  We also examine the dependence of the
r-process yields on the neutron star mass in order to assess the
relation between the stellar mass and the outer boundary condition.
Since a different mass of progenitor star is suggested to lead to a
different mass of envelope and remnant mass, we infer a correspondence
between the progenitor mass and the outer boundary condition to
realize the universality and successful the r-process abundance
pattern.  This study is important to constrain the mass range of
progenitor stars that culminate their evolution for collapse-driven
supernovae as the source of the r-process elements which are 
observed in very metal-poor stars.

\section{Models of Proto-Neutron Star and Neutrino-Driven Winds}

In collapse-driven supernovae, the shockwave launches at the bounce of
central part of collapsing iron core and a proto-neutron star forms
after the core bounce.  A hot and less-dense region, which is often
called a hot bubble, above the proto-neutron star behind the shockwave
is believed to be an ideal site for the r-process.  Thin surface layer
of the proto-neutron star is heated by intense flux of neutrinos,
increasing the entropy and forming a neutrino-driven wind as a hot and
less-dense gas. Thus, the wind is finally blown off toward the
shockwave from behind.  At this time the shockwave has already been
propagating outward further at $\sim 10^{3} - 10^{4}$ km.

We perform numerical simulations of the hydrodynamics of the thin
layer of the proto-neutron star, which is parameterized by mass and
radius, to obtain the temporal information of thermodynamical
conditions for nucleosynthesis.  The details of the numerical
simulation can be found in Sumiyoshi et al. (2001).  We employ the
implicit lagrangian code for general relativistic and spherically
symmetric hydrodynamics (Yamada 1997) including the heating and
cooling processes due to neutrinos (Qian and Woosley 1996).  The
average energy of neutrinos is set equal to $10$ MeV for electron-type
neutrinos, $20$ MeV for electron-type anti-neutrinos, and $30$ MeV for
$\mu$- and $\tau$-neutrinos and their anti-neutrinos, as adopted in
the previous studies (Qian and Woosley 1996, Otsuki et al. 2000,
Sumiyoshi et al. 2000). We set the neutrino luminosity $L_{\nu_i} =
1 \times 10^{51}$ erg/s for each species ($i =$ e, $\nu$, $\tau$) and
it is set to be constant during the simulation.  Accordingly, the
total luminosity is $L_{\nu}^{total} = 6 \times 10^{51}$ erg/s.  The
neutrino distribution function is set at each lagrangian grid point
instead of solving the Boltzmann equation (Yamada, Janka $\&$ Suzuki
1999).  We use the extended table of the relativistic equation of
state (Shen et al. 1998a, 1998b, Sumiyoshi et al. 2001).

We adopt a neutron star with the typical mass, $M_{NS} = 1.4 M_{\odot}$,
and the radius, $R_{NS} = 10$ km, as an inner boundary condition.  We
obtain the initial structure of thin layer above this neutron star by
solving the Oppenheimer-Volkoff equation of the hydrostatic equilibrium.
The density range of this layer is from $10^{11}$ to $5 \times 10^{6}$
g/cm$^3$ and the covered baryon mass is about $10^{-5} M_{\odot}$.  This
means that surface material is thin enough and the assumption of
constant temperature and constant electron fraction holds well.  For
simplicity, we set the temperature to $3$ MeV and the electron fraction
to $0.25$ in the whole material as typical values. The results  are not
sensitive to these values.

As an outer boundary condition, we put a constant pressure next to the
outermost grid point of lagrangian mesh. We parameterize this pressure
of a uniform gas as the outer boundary pressure, $P_{out}$ and study
its effect on the neutrino-driven wind and a subsequent r-process.
This pressure does not influence the wind near the surface of a
neutron star, because $P_{out}$ is much smaller than the pressure of
the whole layer at high density.  It acts as a deceleration of the
expansion only when the pressure of the material becomes close to the
$P_{out}$ value. The pressure approaches this value asymptotically as
the wind material goes far away from a neutron star.  Accordingly, the
temperature of wind material also approaches a final temperature
$T_{out}$, corresponding to $P_{out}$, since the pressure is dominated
by radiation.  A choice of the outer boundary pressure has apparently
a big influence on the r-process nucleosynthesis because the r-process
takes place most efficiently when the wind material arrives at the
region just behind the shockwave (Terasawa et al. 2001). This fact
demands fairly careful studies of the outer boundary conditions in
supernova explosions.

In most previous studies of neutrino-driven winds, the boundary
condition of temperature has been uniquely chosen to $0.1$ MeV at
radius $10^{4}$ km according to the pioneering study of Woosley et
al. (1994).  The value of outer boundary pressure, $P_{out}$, was
taken to be $10^{22}$ dyn/cm$^2$ in the previous hydrodynamical study
of the neutrino-driven wind (Sumiyoshi et al. 2001) to match with the
temperature of about $0.1$ MeV.  The $P_{out}$ value, however, should
change in a certain allowed range because the thermodynamical
conditions of the region between the neutron star surface and the
shockwave may depend on each progenitor star.  This means that a
higher $P_{out}$ value might correspond to a more massive star
associated with more massive envelope and a larger iron core, which
may be related with both neutron star mass and outer boundary
pressure.  In principle it should be determined by the configuration
of progenitor and the passage of the shockwave. It is, therefore,
necessary to carry out the numerical simulations of the whole dynamic
processes from core-collapse to explosion in order to determine the
conditions precisely. Analytical studies of shock propagation are also
important in order to evaluate the pressure value behind the shock
wall.

In the current study we vary the outer boundary pressure as $P_{out} =
10^{20}$, $10^{21}$, $10^{22}$ dyn/cm$^2$ so as to cover reasonable
range for various supernova environments instead of carrying out
numerical simulations of the whole dynamics.  These pressure
correspond to final temperatures ranging $T_9 \sim 0.4 - 1.3$.  Note
that a higher pressure such as $10^{23}$ dyn/cm$^2$ is not used since
only the iron group elements are finally synthesized by dominating
alpha-capture reactions due to the high temperature $0.1$ MeV $ <
T_{out}$, which is irrelevant in the present studies of the r-process
nucleosynthesis.

We perform a set of numerical simulations for different $P_{out}$
values and take out one of trajectories from the lagrangian mesh in
each simulation for nucleosynthesis calculations.  We summarize in
Table 1 the key quantities for the r-process nucleosynthesis from the
numerical results of hydrodynamical simulations.  In this table
$\tau_{dyn}$, $S$, $Y_{e,i}$ and $T_{out}$ stand for the dynamical
timescale, the entropy per baryon (in the unit of the Boltzmann
constant), the initial electron fraction for nucleosynthesis when the
temperature drops to $T_9 = 9.0$, and the asymptotic temperature (in
the unit of $10^9$ K) of the expanding wind material.  We call this
asymptotic temperature as the outer boundary temperature since the
temperature approaches this value to be determined by the outer
boundary pressure.  Note that the definition of $\tau_{dyn}$ is the
$e$-fold time at $T = 0.5$ MeV, which is conventionally used to
characterize the duration of the alpha-process.  Let us emphasize that
the expansion timescales in our model calculations are a few times
larger than those in the previous models, $\tau_{dyn} \le 10$ ms,
which lead to a successful r-process abundance pattern (Otsuki et
al. 2001, Sumiyoshi et al. 2001).  Nevertheless, the expansions in all
calculated models are rapid enough so that the r-process occurs
promptly even at asymptotic region close to the shock wall without
hindrance by the neutrino-process (Meyer 1995).

We show the time evolution of the temperature, $T_9$, in the upper
panel of Fig. \ref{ntos}.  The dashed, solid, dotted lines are the
results in the cases of $P_{out} = 10^{20}$, $10^{21}$, $10^{22}$
dyn/cm$^2$, respectively.  Note that the dynamics of the winds is
similar to one another among three cases at the temperature $T_{9} \ge
7$, below which a gradual departure towards the final stage of
different outer boundary conditions of $P_{out}$ and $T_{out}$ can be
seen.  The time evolutions of the abundances of alpha particles and
seed nuclei are different among three cases, as shown in
Fig. \ref{ntos}, due to the different temperature variation below
$T_{9}=7$.  These differences lead to a significant difference in the
products of the r-process nucleosynthesis.  In Table 1,
$Y_{\alpha,out}$ and $Y_{seed,out}$ are the final abundances of alpha
particles and the sum of seed nuclei ($70 \le A \le 120$) at the final
stage of our nucleosynthesis calculations.  The details of these
quantities are discussed in the next section.

\section{Results of the Nucleosynthesis calculations}

We employ the reaction network including over 3000 species from the
$\beta$-stability line to the neutron drip line including light
neutron-rich unstable nuclei (Terasawa et al. 2001), which are
connected by more than 10,000 nuclear reactions and the
neutrino-processes.  In this paper we use one trajectory corresponding
to the same mass region in each model calculation and adopted the time
evolutions of $T$ and $\rho$ in order to calculate the r-process
abundances.  Note that we can get nearly the same abundance pattern
for all trajectories except for the edge of calculated region. We
start the network calculations of the nucleosynthesis at the time when
the temperature drops to $T_9 = 9.0$, which refers to the time t $=
0.0$ s, and follow the time evolution of the abundances.

\begin{figure*}
\includegraphics[width=\textwidth]{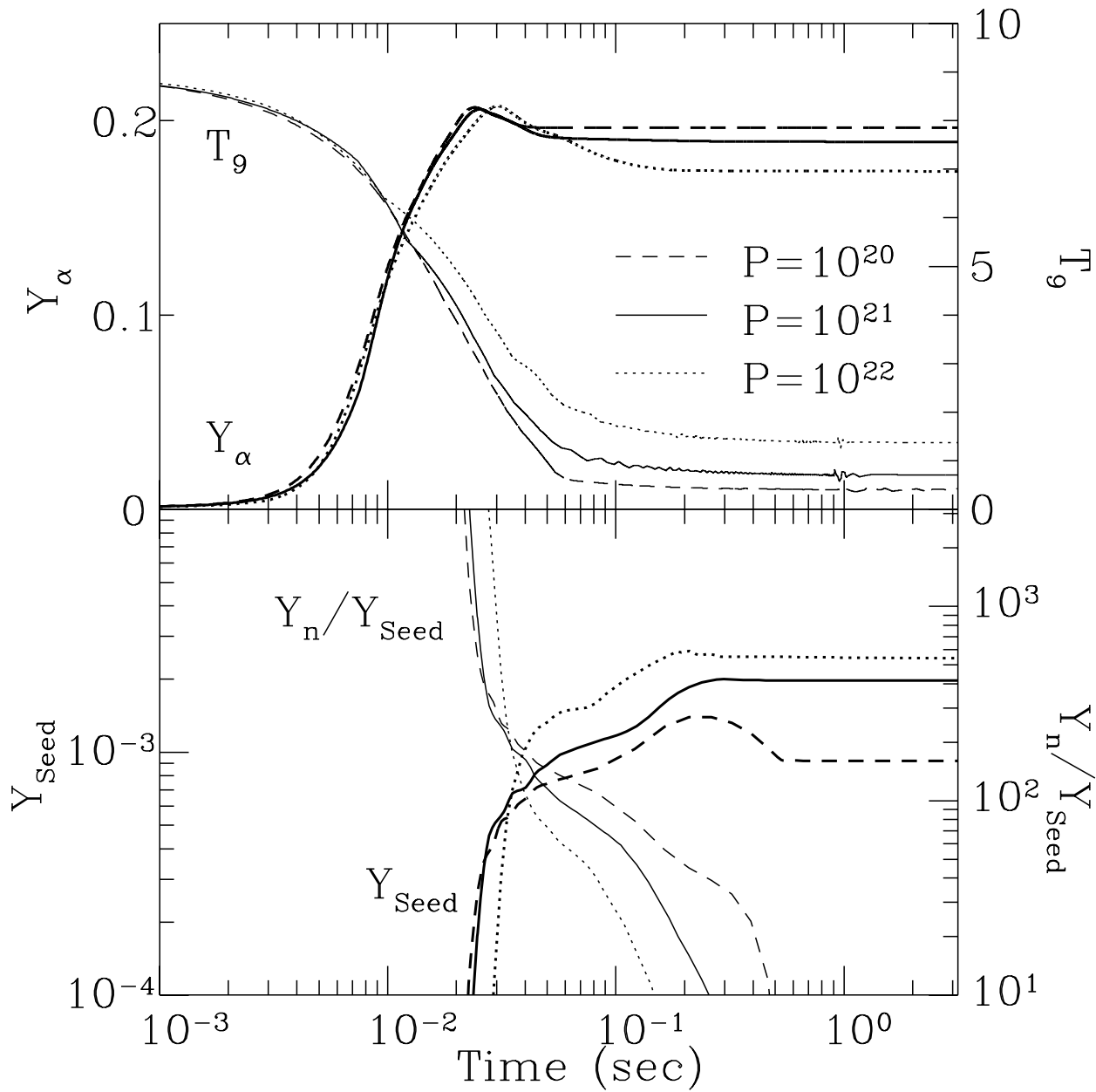} 
\caption{Time variations of the alpha-particle abundance, 
$Y_{\alpha}$, and temperature $T_9$ (upper panel), and neutron-to-seed
ratio, $Y_n/Y_{seed}$, and seed abundance,$Y_{seed}$ (lower
panel). The dashed, solid, dotted lines are the results in the cases
of $P_{out} = 10^{20}$, $10^{21}$, and $10^{22}$ dyn/cm$^2$, respectively.
\label{ntos}}
\end{figure*}

\begin{figure*}
%\epsscale{2.}
%%%\plotone{final.eps}
\includegraphics[width=\textwidth]{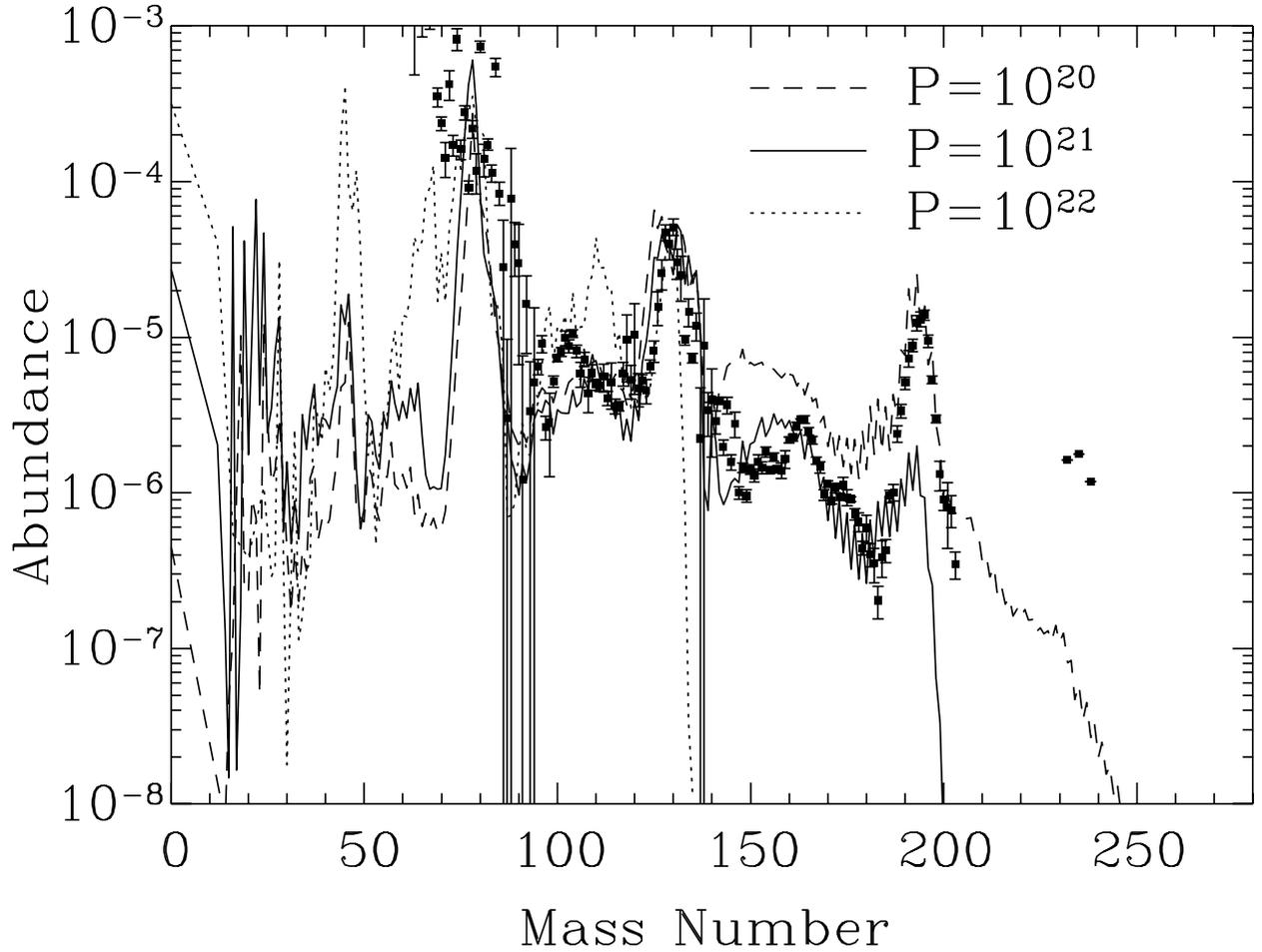} 
\caption{Final abundance yields as a function of the mass number.
The dashed, solid, dotted lines are the same as those in Fig. 1. Data
points are the solar r-process abundances in arbitrary unit from
K$\ddot{\rm a}$ppeler et al. (1989).
\label{final}}
\end{figure*}

We show in Fig. \ref{final} the calculated final abundance patterns as
a function of mass number.  The dashed, solid, and dotted lines are
the results in the cases of $P_{out} = 10^{20}$, $10^{21}$, and
$10^{22}$ dyn/cm$^2$, respectively, as in Fig. \ref{ntos}.  For
comparison the solar system r-process abundance pattern (K$\ddot{\rm
a}$ppeler et al. 1989) is shown by points in arbitrary unit. When we
adopt the outer boundary pressure of $P_{out} = 10^{20}$ dyn/cm$^2$,
the characteristic three peaks in the r-process abundance pattern are
well reproduced at around A$ \sim 80, 130$ and $195$. On the other
hand, in the case of $P_{out} = 10^{21}$ dyn/cm$^2$, 3-rd peak
elements are underabundant.  In the highest pressure model with
$P_{out} = 10^{22}$ dyn/cm$^2$, the nuclear reaction flow stops at the
region of only the 2-nd peak, and the 3-rd peak elements are not
formed at all.  From this figure we can see that more abundant heavy
elements are synthesized as $P_{out}$ is lower.  Since there is a
clear and reasonable relation between $T_{out}$ and $P_{out}$ (Table 1
and Fig. \ref{ntos}), we can conclude that heavier r-process elements
are more abundantly synthesized when $T_{out}$ is lower.  We will
describe the reasons for this later.

Let us discuss the mechanism of the nucleosynthesis quantitatively
more in detail. The r-process nuclei are synthesized by two
sequential processes, alpha-process and r-process (Woosley and
Hoffman 1992).  Near the surface of neutron star the matter maintains
the nuclear statistical equilibrium (NSE) and there are only free
neutrons and protons because of the high temperature ($T_9 \sim
30$). When the temperature becomes lower than $T_9 \sim 10$, free
neutrons and protons begin to assemble into composite nuclei.

In Fig. \ref{ntos}, the abundance of alpha particles, $Y_{\alpha}$, in
the upper panel, and the abundance of seed nuclei, $Y_{seed}$, and the
neutron-to-seed ratio, $Y_n/Y_{seed}$, in the lower panel are shown as
a function of time when the temperature becomes $T_9 < 9$. The value
of $Y_{\alpha}$, at first, rapidly increases in the NSE. As the
temperature decreases, the system is gradually out of NSE, and
$Y_{seed}$ starts increasing when $Y_{\alpha}$ approaches the peak
abundance around $T_9 \sim 5$. After this time, alpha particles are
gradually consumed by the alpha-process in order to produce
neutron-rich seed nuclei with mass number A$\sim 100$.  The peak
position of $Y_{\alpha}$ is different in the three models for
different outer boundary pressures because of the difference in the
dynamical timescale.  Note, however, that this difference in the
timescale makes a smaller influence on the nucleosynthesis than the
influence arising from different $T_{out}$. As the temperature becomes
lower, charged particle reactions become rapidly slower, and
alpha-captures also cease when the temperature becomes close to $T_9 <
2$ towards the alpha-rich freezeout. At this time and temperature the
composition changes to neutrons, alphas, and a little amount of seed
nuclei. The r-process starts from these seed nuclei and makes heavy
r-process elements. Heavy elements are synthesized more abundantly, as
the number of neutrons per seed nucleus, $Y_n/Y_{seed}$, becomes
larger.  Finally, neutrons are almost consumed and $Y_{seed}$
eventually reaches almost constant value. At this freezeout time of
the r-process, the nuclear flow almost ceases and produced
neutron-rich nuclei begin to convert into stable nuclei, smoothing the
abundance pattern by $\beta$-decays and $\beta$-delayed neutron
emissions.

The $Y_n$ values are almost the same in our three models because
entropies are almost the same (Table 1). The important factor to make
the difference in final abundance pattern is the abundance of seed
nuclei. In the model with a short dynamical timescale for $P_{out} =
10^{20}$ dyn/cm$^2$ and $T_{out} \sim 0.4$, since the temperature
drops rapidly (see the dashed line in upper panel of Fig. \ref{ntos}),
the alpha-process does not proceed efficiently as it does in the slow
expansion model like Woosley's flow (Woosley et al. 1994). This is
because charged particle reactions occur at high temperatures. There
are relatively small amount of seed nuclei produced near the boundary
region where the temperature drops to $T_{out}$.  This scenario,
however, tacitly assumes that the temperature is so low as not to
operate the alpha-process efficiently. If, on the other hand, the
temperature is high ($T_{out} \geq 1.0$) in the outer boundary
region, alpha captures can frequently occur to the same or even
greater extent as neutron captures. The seed nuclei continue to be
made by the alpha-process in such cases.

This mechanism makes the biggest influence on the seed abundance and
the r-process nucleosynthesis as well. When $T_{out}$ is high,
Y$_{\alpha,out}$ is small and Y$_{seed,out}$ is large (Table 1 and
Fig. \ref{ntos}).  As a result, in the high pressure model,
$Y_n/Y_{seed}$ is smaller than the other low pressure models, and the
heavy elements cannot be synthesized, as shown in Fig. 2. In the case
of the low pressure model, on the contrary, the temperature is as low
as $T_9 \sim 0.4$ in the outer boundary region. Therefore, even after
the alpha-rich freezeout there are plenty of free neutrons with
relatively smaller amount of seed elements, because the alpha captures
are suppressed due to the low temperature, and the r-process proceeds
creating heavy elements at the 3rd r-process peak abundantly (the
dashed curve in Fig. 2).

\section{Discussions and Conclusion}

We showed a possibility that the r-process can successfully occur in
the neutrino-driven winds from a neutron star having typical observed
neutron star mass, $1.4 M_{\odot}$, provided that the outer boundary
condition is appropriately chosen.  More specifically from Fig. 1, we
can conclude that the model with $P_{out} = 10^{20}$ dyn/cm$^2$
(dashed line) is most likely in the neutron star model with $M_{NS}
=1.4 M_{\odot}$ and $R_{NS} = 10$ km.  In the present model the
dynamical timescale is a few times longer than the previous successful
models and the entropy per baryon is relatively low, $100-200$ k$_{\rm
B}$.

This conclusion results from the fact that the alpha captures do not
efficiently work as the outer boundary temperature, $T_{out}$, becomes
lower.  We can reconfirm this effect quantitatively in the following way.  If the
seed nuclei are synthesized only by the alpha-process and the seeds do
not change to other heavy nuclei approximately, the decrease in alpha particles from the peak value in Fig. 1 is equal to
the increase in the seeds. Then, the relation, $\Delta Y_{\alpha}
\times 4 = \Delta Y_{seed} \times 100$, follows due to the mass
conservation, where we assume the averaged mass number of seed nuclei
100.  We could confirm actually that this relation holds very well in
final abundances ($Y_{seed,out}$ and $Y_{\alpha,out}$) in our three
models (Table 1).

It is generally believed that the neutron star mass depends somewhat
on the progenitor mass (Woosley \& Weaver 1995, Thielemann et
al. 1996, Timmes et al. 1996, Limongi et al. 2000).  The physical
conditions that govern the r-process are determined by $S$, $Y_e$, and
$\tau_{dyn}$, which also strongly depend on the neutron star
mass. However, recent observations of neutron-capture elements in
metal-deficient stars show the universality in the abundance pattern
for the r-process elements with $56
\le Z \le 70$ (Sneden et al. 2000, Westin et al. 2000, Cayrel et
al. 2001, Honda 2002). They suggest that the universal r-process
abundance pattern should be realized independent of the neuron star
mass.  Let us consider the dependence of abundance pattern on the
neutron star mass by using the same outer boundary pressures.  We
adopt several neutron star mass of $1.2$, $1.3$, $1.4$, $1.5$ and $1.6
M_{\odot}$ in the simulations of the neutrino-driven winds (Terasawa
2002 and Terasawa et al. 2002). These masses cover the almost all
observed neutron star masses except for a few neutron stars (Bulik et
al. 1995, Brown et al. 1996, Thorsett and Chakrabarty 1999, Barziv et
al. 2001, and references therein).

As for the outer boundary conditions, we varied the pressure values as
$P_{out} = 10^{20}$, $10^{21}$, and $10^{22}$ dyn/cm$^2$ in our
simulations of the $1.2 - 1.6 M_{\odot}$ models. The value of $S$ is
higher and $\tau_{dyn}$ is shorter as the neutron star mass becomes
larger, although $Y_{e,i}$ are almost the same $\sim 0.43 - 0.44$ in
all calculations, which we have carried out, and $T_{out}$ is common
for each value of $P_{out}$. As discussed in the previous section, we
saw that $Y_{\alpha,out}$ becomes smaller and $Y_{seed,out}$ becomes
larger with increasing pressure for the fixed neutron star mass. On
the other hand, for the fixed outer boundary pressure,
$Y_{\alpha,out}$ becomes larger and $Y_{seed,out}$ becomes smaller as
the neutron star mass becomes larger. From these systematics which we
found, we understand that more abundant heavy elements are synthesized
as the neutron star mass is larger and the outer boundary pressure
becomes lower.  In the previous study of the r-process nucleosynthesis
in neutrino-driven winds (Sumiyoshi et al. 2001), the outer boundary
pressure was set equal to $P_{out} \sim 10^{22}$ dyn/cm$^2$ for both
neutron star mass models of $1.4 M_{\odot}$ and $ 2.0 M_{\odot}$.
This pressure ($P_{out} \sim 10^{22}$ dyn/cm$^2$) corresponds to the
temperature $T_{out} \sim 0.1$ MeV.  By their calculations, when this
value was adopted in the $1.4 M_{\odot}$ model, the flow of
nucleosynthesis virtually stopped at the nuclear mass region below the
2-nd peak.  Therefore, a higher neutron star mass model ($2.0
M_{\odot}$) was adopted to increase the entropy for a successful
r-process. This theoretical correlation among $P_{out}$, $T_{out}$,
and $M_{NS}$ is reasonably understood, that is a higher outer boundary
pressure corresponds to a higher neutron star mass, and vice
versa. However, it is to be stressed that increasing boundary pressure
and temperature leads to unsuccessful r-process without increasing the
neutron star mass.  In the present study we find another condition to
realize successful r-process namely by adopting suitable outer
boundary condition with keeping the neutron star properties as those
measured observationally, i.e. $M_{NS} \sim 1.4 M_{\odot}$ and $R_{NS}
\sim 10$ km.  This result makes a strong constraints on modeling the
dynamics of supernova explosion in view of constructing the structure
model of massive progenitor stars in order to clarify the physical
conditions of the r-process nucleosynthesis.

\begin{deluxetable}{ccccccc}
\tabletypesize{\scriptsize}
\tablecaption{The Key Quantities for the \lowercase{r}-Process Nucleosynthesis.  \label{tbl-1}}
\tablewidth{0pt}
\tablehead{
\colhead{P$_{out}$ [dyn/cm$^2$]} & \colhead{$\tau_{dyn}$ [sec]}   & \colhead{ S [k$_B$]}   &
\colhead{$Y_{e,i}$} &
\colhead{T$_{out}$ [$10^9$K]}  & \colhead{$Y_{\alpha,out}$} & \colhead{$Y_{seed,out}$}}
\startdata

 $10^{20}$   &  $2.32 \times 10^{-2}$   & $200$ &
 0.43 & 0.4 &$0.196$&$9.2 \times 10^{-4}$\\
 $10^{21}$   &  $2.54 \times 10^{-2}$   & $180$ &
 0.43 & 0.7 &$0.189$&$2.0 \times 10^{-3}$\\
 $10^{22}$   &  $3.34 \times 10^{-2}$   & $170$ &
 0.44 & 1.3 &$0.174$&$2.4 \times 10^{-3}$\\  

 \enddata

%% Text for table notes should follow after the \enddata but before
%% the \end{deluxetable}. Make sure there is at least one \tablenotemark
%% in the table for each \tablenotetext.

\tablecomments{The summary of model parameters and key quantities for the
 r-process. P$_{out}$ is the pressure at the outer boundary and we give
 these values as an outer boundary condition. $\tau_{dyn}$, $S$,
 and $Y_{e,i}$ are the dynamical timescale, entropy per
 baryon, and initial electron fraction,
 respectively. Note that the definition of $\tau_{dyn}$ is the $e$-fold
 time at $T = 0.5$ MeV. $Y_{\alpha,out}$ and $Y_{seed,out}$ are the
 final abundances of alpha-particles and seed nuclei.}
\end{deluxetable}

\acknowledgments

One of the authors (MT) wishes to acknowledge the fellowship of the
Japan Society for Promotion of Science (JSPS) and would like to thank
Professor K. Ikeda and Dr. N. Itagaki for helpful discussions. This
work is in part supported by JSPS under the Grants-in-Aid Program for
Scientific Research (10640236, 10044103, 11127220, 12047230, 12047233,
13002001, 13640313, 13740165, 14039210) of for the Ministry of
Education, Science, Sports, and Culture of Japan.


\begin{thebibliography}{9}

\bibitem{} 
Barziv, O., Kaper, L., Van Kerkwijk, M. H., Telting, J. H., and Van
 Paradijs, J.,  2001, A \& A, 377, 925 

\bibitem{} 
 Brown, G. E., Weingartner, J. C., Wijers, and Ralph A. M. J., 1996, ApJ, 463, 297  

\bibitem{} 
 Bulik, T., Riffert, H., Meszaros, P., Makishima, K., Mihara, T., and 
 Thomas, B., 1995, ApJ, 444, 405

\bibitem{}
Burbidge, E. Margaret, Burbidge, G. R., Fowler, William A., Hoyle, F. 1957,
Rev. Mod. Phys., 29, 547.

\bibitem{} 
 Cayrel, R., Hill, V., Beers, T. C., Barbuy, B., Spite, M., Spite, F.,
			    Plez, B., Andersen, J., Bonifacio, P.,
			    Francois, P., Molaro, P., Nordstrom, B.,
			    Primas, F., 2001, Nature, 409, 691

\bibitem{} 
Honda, S. et al. (SUBARU/HDS Collaboration), 2002, in preparation

\bibitem{}
 Johnson, J. A., Bolte, M., 2001, Nucl. Phys, A688, 41

\bibitem{} 
K\"appeler, F., Wiescher, M., Giesen, U., Goerres, J., Baraffe, I., El
Eid, M., Raiteri, C. M., Busso, M., Gallino, R., Limongi, M., and
Chieffi, A.,1989, ApJ, 437, 396

\bibitem{}
Limongi, M., Straniero, O., Chieffi, A., 2000, ApJ., 129, 625

\bibitem{2}
Meyer, B. S. 1995, ApJ. Lett., 449, 55

\bibitem[Otsuki et al. 2000]{otauki00}
Otsuki, K., Tagoshi, H., Kajino, T., and Wanajo, S. 2000, \apj, 533, 424

\bibitem[Qian \& Woosley 1996]{Qian96}
   Qian, Y. -Z. and Woosley, S. E. 1996, \apj, 471, 331

\bibitem[Shen et al.~1998a]{Shen98a}
   Shen, H., Toki, H., Oyamatsu, K., and Sumiyoshi, K. 1998a, \nphysa, 637, 435

\bibitem[Shen et al.~1998b]{Shen98b}
Shen, H., Toki, H.,
Oyamatsu, K., and Sumiyoshi, K.  1998b  Prog. Theor. Phys., 100, 1013

\bibitem[Sneden et al.~1998]{Sned98}
   Sneden, C., Cowan, J. J., Debra, L. B., and Truran, J. W. 1998, \apj, 496,
235

\bibitem[Sneden et al.~2000]{Sned00}
Sneden, C., Cowan, J. J., Ivans, I. I. Fuller, G. M.
Burles, S., Beers, T. C., and Lawler, J. E. 2000,
\apjl, 533, L139

\bibitem[Sneden et al.~1996]{Sned96}
   Sneden, C., McWilliam, A., Preston, G. W., Cowan, J. J., Burris, D. L.,
   and Armosky, B. J. 1996, \apj, 467, 819

\bibitem{3}
Sumiyoshi, k., Suzuki, H., Otsuki, K., Terasawa, M.,
  and  Yamada, S. 2000, Pub. Astron, Soc. Japan  52, 601 

\bibitem{4}
Terasawa, M., Sumiyoshi, K., Kajino, T., Tanihata, I.,
  and Mathews, G. J. 2001, ApJ, Vol. 562, pp. 470-479

\bibitem{5}
Terasawa, M. 2002, PhD Thesis, University of Tokyo 

\bibitem{5}
Terasawa, M., Sumiyoshi, K., Yamada, S., Suzuki, H. 2002, in preparation

\bibitem{}
Thielemann, F. K., Nomoto, K., and Hashimoto, M. 1996, ApJ, 460, 408

\bibitem{}
Timmes, F. X., Woosley, S. E., Weaver, Thomas A., 1996, ApJ, 457, 834

\bibitem{} 
 Thorsett, S. E., and  Chakrabarty, D., 1999, ApJ, 512,288  

\bibitem{}
Westin, J.,  Sneden, C., Gustafsson, B., and  Cowan, J. J. 2000, ApJ,
			    530, 783

\bibitem[Woosley \& Hoffman 1992]{Woos92}
   Woosley, S. E. and Hoffman, R. D. 1992, \apj, 395, 202

\bibitem[Woosley \& Weaver 1995]{WW95}
Woosley, S.E. and Weaver, T.A. 1995, ApJS, 101, 181

\bibitem{1}
Woosley, S. E., Wilson, J. R.,  Mathews,G. J., Hoffman, R. D., and
Meyer, B. S. 1994, ApJ, 433, 229

\bibitem[Yamada 1997]{Yam97}
Yamada, S. 1997, ApJ,  475, 720

\bibitem[Yamada et al. 1999]{Yam99}
Yamada, S. Janka, H.-Th., and Suzuki, H. 1999, A\&A, 344, 533

\end{thebibliography}
\end{document}